\documentclass{aa}  

\usepackage{graphicx}
\usepackage{txfonts}
\usepackage{textcomp}
\usepackage{subdepth}
\begin{document}

\title{Radio emission in ultracool dwarfs: the nearby substellar triple system VHS\,1256$-$1257}


\author{J. C. Guirado\inst{1,2},
R. Azulay\inst{1,3},
B. Gauza\inst{4,5,6},
M. A. P\'erez-Torres\inst{7,8},
R. Rebolo\inst{4,5,9},
J.B. Climent\inst{1}
\and
M.R. Zapatero Osorio\inst{10}
}

\institute{Departament d'Astronomia i Astrof\'{\i}sica, Universitat de Val\`encia, C. Dr. Moliner 50, E-46100 Burjassot, Val\`encia, Spain\\
\email{guirado@uv.es}
\and
Observatori Astron\`omic, Universitat de Val\`encia, Parc Cient\'{\i}fic, C. Catedr\'atico Jos\'e Beltr\'an 2, E-46980 Paterna, Val\`encia, Spain
\and
Max-Planck-Institut f\"ur Radioastronomie, Auf dem H\"ugel 69, D-53121 Bonn, Germany
\and
Instituto de Astrof\'isica de Canarias, E-38200 La Laguna, Tenerife, Spain
\and
Departamento de Astrof\'isica, Universidad de La Laguna, E-38206 La Laguna, Tenerife, Spain
\and
Departamento de Astronom\'ia, Universidad de Chile, Camino el Observatorio 1515, Casilla 36-D, Las Condes, Santiago, Chile
\and
Instituto de Astrof\'isica de Andaluc\'ia (IAA, CSIC), Glorieta de la Astronom\'ia, s/n, E-18008 Granada, Spain
\and
Departamento de F\'isica Te\'orica, Facultad de Ciencias, Universidad de Zaragoza, E-50009 Zaragoza, Spain
\and
Consejo Superior de Investigaciones Cient\'ificas, E-28006 Madrid, Spain
\and
Centro de Astrobiolog\'{\i}a (CSIC-INTA), Crta. Ajalvir km 4, E-28850 Torrej\'on de Ardoz, Madrid, Spain 
}

   \date{Accepted  2017 December 1.}


  \abstract
{}
{With the purpose to investigate the radio emission of new ultracool objects, we carried out a targeted search in the recently discovered system VHS\,J125601.92$-$125723.9 (hereafter VHS\,1256$-$1257); this system is composed by an equal-mass M7.5 binary and a L7 low-mass substellar object located at only 15.8\,pc.}
{We observed in phase-reference mode the system VHS\,1256$-$1257 with the Karl G. Jansky Very Large Array at $X$- and $L$- band and with the European VLBI Network at $L$-band in several epochs during 2015 and 2016.} 
{We discovered radio emission at $X$-band spatially coincident with the equal-mass M7.5 binary with a flux density of 60\,$\mu$Jy. We determined a spectral 
index $\alpha = -1.1 \pm 0.3$ between 8 and 12\,GHz, suggesting that non-thermal, optically-thin, synchrotron or gyrosynchrotron radiation is 
responsible for the observed radio emission. Interestingly, no signal is seen at $L$-band where we set a 
3-$\sigma$ upper limit of 20\,$\mu$Jy. This might be explained by strong variability of the binary or self-absorption at 
this frequency. By adopting the latter scenario and gyrosynchrotron radiation, we constrain the turnover frequency to be 
in the interval 5--8.5 GHz, from which we infer the presence of kG-intense magnetic fields in the M7.5 binary. 
Our data impose a 
3-$\sigma$ upper bound to the radio flux density of the L7 object of 9\,$\mu$Jy at 10\,GHz.}
{}

    
\titlerunning{Radio emission in ultracool dwarfs: the nearby planetary system VHS\,1256$-$1257}

\authorrunning{Guirado et al.}
   \maketitle
%

\begin{table*}
\label{journal}      
\begin{center}
\caption{VLA and VLBI Observations of VHS\,1256--1257}
\resizebox{\hsize}{!}{         
\begin{tabular}{lccccccc}
\hline\hline 
      \noalign{\smallskip}
Telescope / Configuration & Epoch & Frequency band & UT Range & Beam size & P.A.  & rms & Peak \\ 
	                                 &           &                           &                &     & [$^\circ$]~~ & [$\mu$Jy]  & [$\mu$Jy]\\
\noalign{\smallskip}
            \hline
            \noalign{\smallskip}
VLA / BnA & 15 May 2015 & $X$  & 05:10\,-\,06:10 & 0.78$^{\prime\prime}$\,$\times$\,0.45$^{\prime\prime}$ & $-$51 & 3 & ~~60 \\
VLA / B     & 28 Jul 2016   & $L$  & 00:00\,-\,01:00 & 5.66$^{\prime\prime}$\,$\times$\,3.38$^{\prime\prime}$ & $-$12 & 7 & $<$21 \\
            \noalign{\smallskip}  
EVN$^\mathrm{a}$ &  4 Mar 2016 & $L$ &  22:30\,-\,04:30 & 3.1\,$\times$\,2.4\,mas & $-$64 & 22 & $<$66 \\          
~~~"                             & 27 May 2016 & " &  15:30\,-\,23:30 & 11.9\,$\times$\,2.2\,mas & $-$78  & 30 & $<$90 \\  
~~~"                             & 2 Nov 2016 & " & 05:00\,-\,13:00 & 3.2\,$\times$\,2.3\,mas  & $-$73  & 31 & $<$93 \\  
      \noalign{\smallskip}
            \hline
            \noalign{\smallskip}  
\end{tabular}
}
\end{center}
\footnotesize{\textbf{Notes.} $^\mathrm{a}$: European VLBI Network using the following antennas: Jodrell Bank, Westerbork, Effelsberg, Medicina, Noto, Onsala85, Tianma65, Urumqi, Torun, Zelenchukskaya, Hartebeesthoek, Sardinia, Irbene, and DSS63.}
\end{table*}

\section{Introduction}

Radio observations play an important role to understand the processes involved in the 
formation and evolution of stellar and substellar objects. In particular,  radio emission studies
of ultracool objects (late M, L, and T objects; e.g., Matthews 2013; Kao et al. 2016, and references therein) are relevant to probe 
the magnetic activity of these objects and its influence on the formation of disks or planets. 
Moreover, the study of ultracool dwarfs may open a suitable route to the detection of 
radio emission of exoplanets: meanwhile no exoplanet has been yet detected at radio wavelengths, 
an increasing number of ultracool objects (McLean et al. 2012) show substantial 
evidence of radio emission at GHz-frequencies in objects with spectral types as cool as T6.5 
(Kao et el. 2016).
But radio detection of ultracool dwarfs is still a relatively rare phenomenon, 
and new candidates are needed to improve the statistics of active cool objects. One of these targets is the system 
VHS\,J125601.92--125723.9 (hereafter VHS\,1256--1257; Gauza et al. 2015); this system is composed by a 0.1$^{\prime\prime}$ equal-magnitude  M7.5 binary (components A and B; Stone et al. 2016) and a lower-mass L7 companion (component b) separated 8$^{\prime\prime}$ from the primary pair. The system is relatively young, 150--300\,Myr, and nearby (12.7--17.1\,pc), which locate the low-mass object b near the deuterium burning limit (Stone et al. 2016). 
Recently, Zapatero Osorio et al. (2017, in preparation) have determined a new trigonometric distance of
15.8$^{+1.0}_{-0.8}$\, pc for the system using optical and near-infrared images spanning a few years. 
This distance is compatible with 
a likely age of 300 Myr, which agrees with the strong lithium depletion observed in the high resolution spectra of the 
M7.5 binary and the recent age determinations of Stone et al. (2016) and Rich et al. (2016). In this work, we will adopt these values of distance and age. 
Using the bolometric corrections available for M7—M8 and red L dwarfs (Golimowski et al. 2004; Filippazzo et al. 
2015) and the new distance, the luminosities are determined at log\,$L/L_\odot$ = $-4.91 \pm 0.10$ dex for the faint L7 companion VHS\,1256$-$1257\,b and log\,$L/L_\odot$ = $-3.24 \pm 0.10$ dex for each member of the M7.5 binary. By applying the luminosity—mass—age relationship of Chabrier et al. (2000), we infer masses in the interval 10–20 M$_{\rm Jup}$  (VHS\,1256$-$1257\,b) and 50–80 M$_{\rm Jup}$ (A and B components) for the age range 150–300 Myr, with most likely values of 15–20 M$_{\rm Jup}$ and 70–80 M$_{\rm Jup}$  at 300 Myr.
The interest in VHS\,1256--1257 is obvious for several reasons: first, it is only the third multiple system 
in which all three components may be substellar (Bouy et al. 2005; Radigan et al. 2013); 
second, the L7 source 
belongs to one intriguing (not yet understood) population of very red L dwarfs with likely high content of atmospheric dust 
or high metallicity (Rich et al. 2016);
third, given 
their large separation (8$^{\prime\prime}$), unambiguous observations of the substellar object b and the central pair AB are accessible by 
instruments at virtually all wavelengths, including radio; and fourth,  
the binarity of the host system AB will permit the determination of their dynamical masses in a few years, which is essential to fully characterize the system.
Additionally, Gauza et al. (2015) reports detection of the H$_\alpha$ line emission (656.3 nm) in the primary, which indicates the existence of chromospheric activity in this M7.5 low-mass binary, therefore showing ability to sustain significant magnetic fields, 
and hence, radio emission.

In this paper we present Karl G. Jansky Very Large Array (VLA) and  European VLBI Network (EVN) observations of VHS\,1256$-$1257. We describe our observations and report the principal results, consisting in the discovery of the radio emission of the central components of the VHS\,1256$-$1257 system. We also present a study of the spectral behaviour of the detected emission and set an upper bound to the possible radio emission of the very low-mass companion VHS\,1256--1257\,b.

\section{Observations and data reduction}

\subsection{VLA observations}
We observed with DDT/Exploratory Time with the VLA the system VHS\,1256$-$1257 at $X$- and $L$-band on 2015 May 15 and 2016 Jul 28, respectively. The observation at $X$-band lasted 2 hours and was carried out in BnA configuration, using an effective bandwidth of 4\,GHz (8$-$12\,GHz) in dual polarization. The observation at $L$-band lasted 1 hour, in B configuration, and using an effective bandwidth of 1\,GHz (1$-$2\,GHz) in dual polarization (see Table 1).
We used 3C286 as absolute flux calibrator meanwhile we performed amplitude and phase calibration using interleaved observations of the radio source J1254$-$1317. Data reduction and imaging were carried out using the CASA\footnote{http://casa.nrao.edu} software package of the NRAO.  The standard procedure of calibration for continuum VLA data was applied. Special care was taken to flag data contaminated by radio frequency interferences (RFI) at $L$-band. The resulting images are shown in Fig. 1.

\subsection{VLBI observations}
The VLA observations explained above confirmed the radio emission of VHS\,1256$-$1257. This detection triggered VLBI observations 
that were carried out with the EVN at $L$-band (1.6\,GHz; see Table 1) with the purpose of constraining both the origin and properties of the radio emission.
Each observation lasted 6--7\,hr and both polarizations were recorded with a rate of 1024\,Mbps (two polarizations, eight subbands per polarization, 16\,MHz per subband, two bits per sample). We used the phase-reference mode and the selected calibrators were J1254$-$1317 (as primary calibrator, separated 0.5\textdegree\,from VHS\,1256$-$1257), and J1303$-$1051 (as secondary calibrator, separated 2.7\textdegree). The duty cycle was 1\,min on the primary calibrator, 3\,min on the target, and 1\,min on the secondary calibrator, with a total integration time at the target of $\sim$3.5\,hr per epoch.

The data reduction was realized using the program Astronomical Image Processing System (AIPS) of the National Radio Astronomy Observatory (NRAO) with standard routines. Once the final data were obtained, the images were made with the Caltech imaging program DIFMAP (Shepherd et al. 1994). We did not detect neither the central M7.5 binary nor the very-low mass substellar companion at any epoch, establishing an average flux density upper limit of  $\sim$80\,$\mu$Jy (3$\sigma$). The interpretation of these non-detection will be discussed in next Section.

\begin{figure*}
\begin{center}
\includegraphics[trim={1cm 3cm 3cm 2cm},clip, width=9.1cm, angle=0]{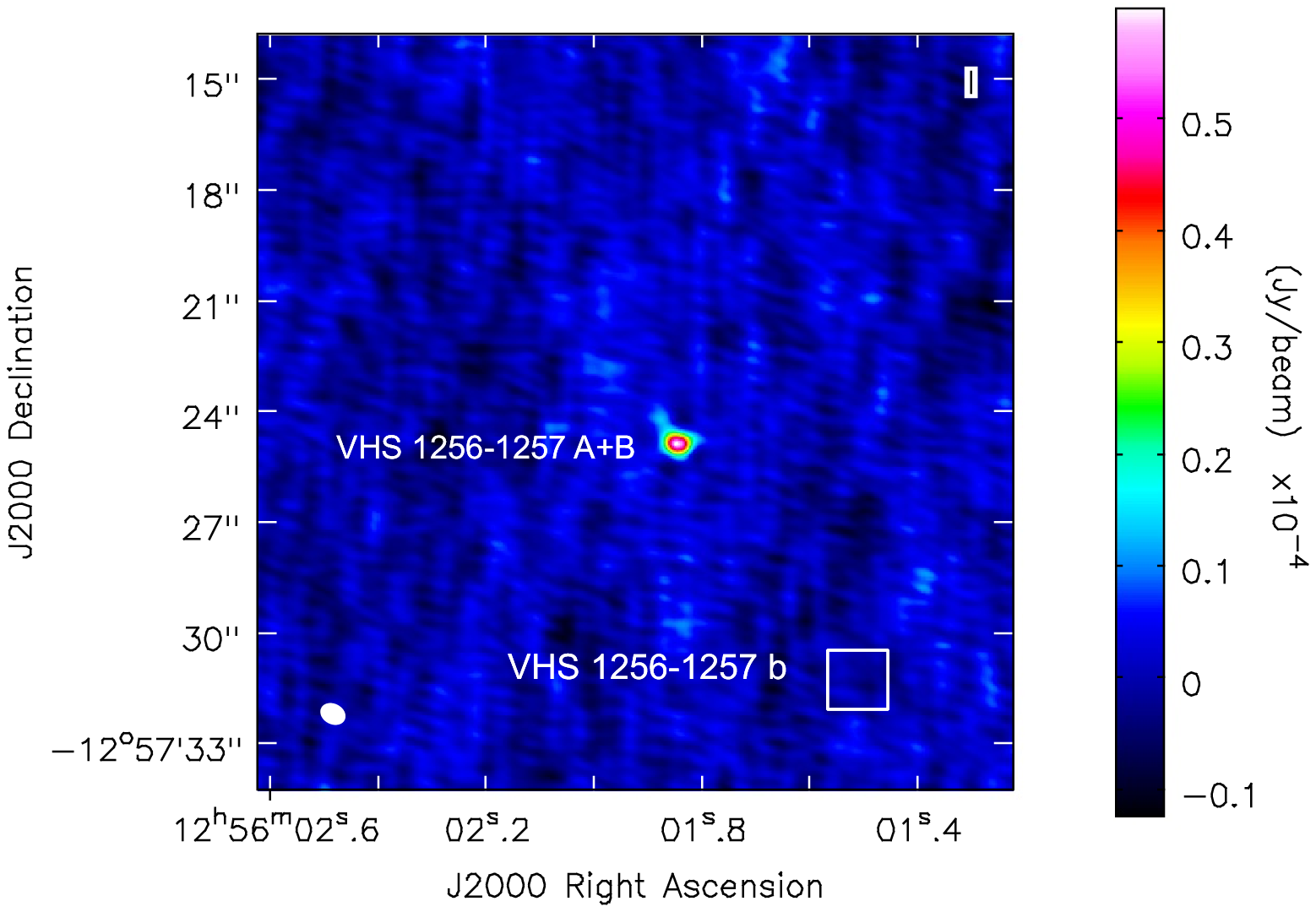}
\includegraphics[trim={1cm 3cm 3cm 2cm},clip, width=9.1cm, angle=0]{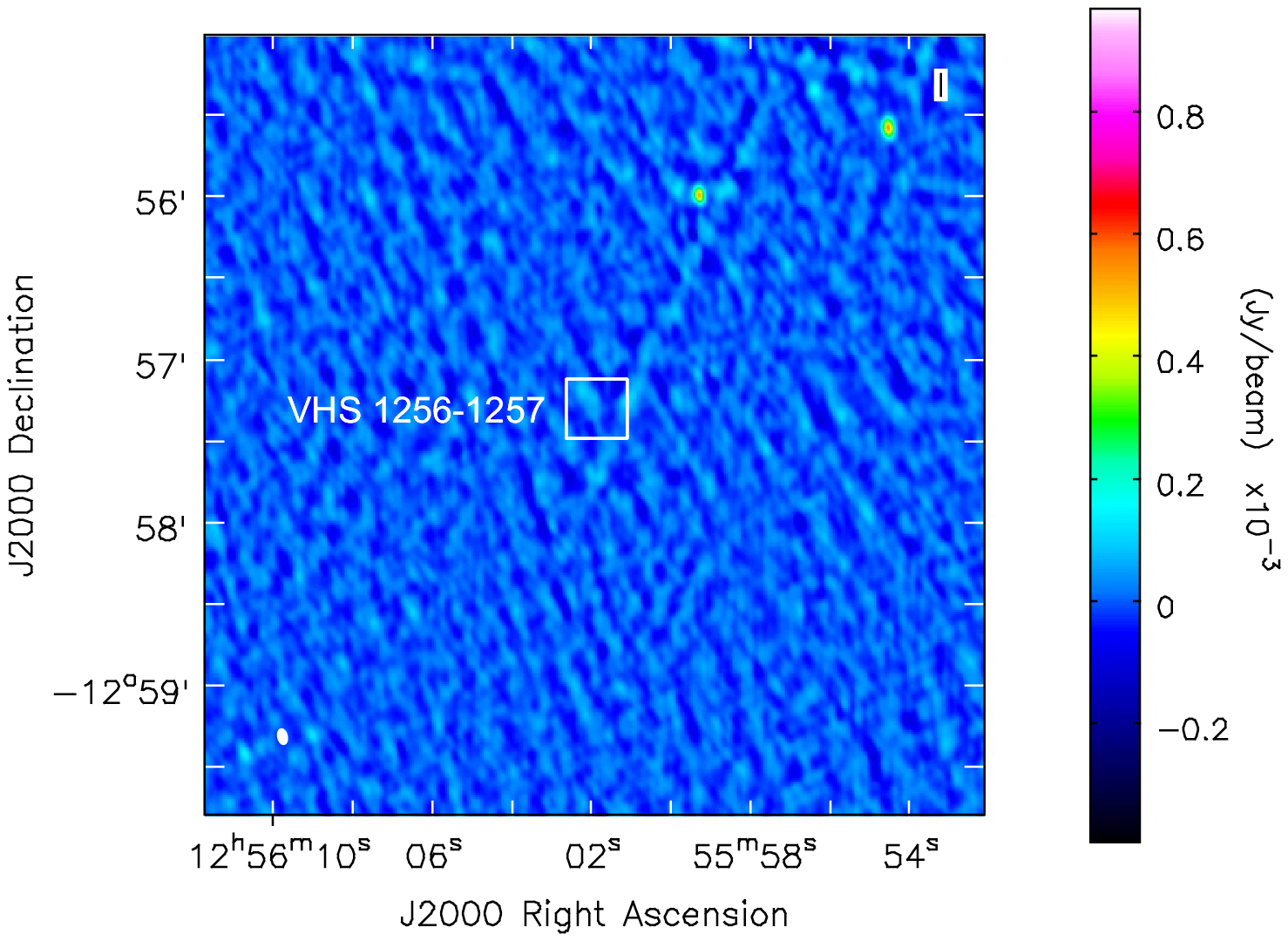}
\vspace{-5.0cm}
\caption{{\it Left.} VLA image of the VHS\,1256$-$1257 field at $X$-band. The detected source is readily assigned to the M7.5 binary VHS\,1256$-$1257\,AB. The location of the (undetected) L7-object b is marked with a solid white box. The 3\,$\sigma$ threshold detection is 9\,$\mu$Jy.  The restoring beam, shown at the bottom left corner, is an elliptical Gaussian of 0.78\,$\times$\,0.45 arcsec (P.A. $-$51$^\circ$). At 15.8\,pc, the separation between components AB and b corresponds to 128.4\,AU.
{\it Right.}  VLA image of the VHS\,1256$-$1257 field at $L$-band. A solid box, with size that of the X-band image, 
is centered at the position of the $X$-band detection. None of the VHS\,1256$-$1257 components is detected at this frequency band. The 3\,$\sigma$ threshold detection is 20\,$\mu$Jy. The two bright knots seen in the map at the NW correspond to known extragalactic radio sources. The restoring beam, shown at the bottom left corner, is an elliptical Gaussian of 5.66\,$\times$\,3.38 arcsec (P.A. $-$12$^\circ$).}
\label{vhsvlax}
\end{center}
\end{figure*}

\section{Results and Discussion}

\subsection{The radio emission of the central pair VHS\,1256--1257\,AB}
Figure \ref{vhsvlax} revealed a clear detection on 2015 May 15 ($X$-band) of an unresolved source with a peak flux density of 60\,$\mu$Jy, which can be assigned to the primary of VHS\,1256$-$1257, the equal-mass M7.5 binary. We confirmed 
this identification by using both the coordinates and proper motion given in Gauza et al. (2015) to find the expected position  
of component AB at the time of our observation; this expected position differs only 0.18$^{\prime\prime}$, 
about one third of the synthesized beam, from the measured position in the VLA $X$-band map in Fig. 1 
(the source has moved  $\sim$6.3$^{\prime\prime}$ between Gauza's epoch and ours).  A noise floor of $\sim$3\,$\mu$Jy imposes a strong upper bound to the radio emission at the expected position of the low-mass companion 
VHS\,1256--1257\,b. In contrast, we found no detection in any of the components of the system on 2016 July 28 ($L$-band), with a 3$\sigma$ threshold detection of 20\,$\mu$Jy. The flux density measured at $X$-band implies a radio luminosity of 
1.95$\times$10$^{-13}$ erg\,s$^{-1}$\,Hz$^{-1}$ at 15.8\,pc. Assuming that the flux is originated at only one of the central components of VHS\,1256--1257, this luminosity is similar to other single ultracool dwarfs detected with comparable spectral types (M7.5; McLean et al. 2012).  We notice that the figures above are halved if we consider both components to contribute equally to the radio flux. 
We did not detect significant traces of variability or pulsed emission in the flux density throughout the 2\,hr duration of our observations, which suggests that the 
detected radio emission is produced either in quiescent conditions or, alternatively, during a long-duration, energetic flare. However, the latter possibility seems unlikely given the low frequency rate of energetic flares in late M dwarfs ($\sim$0.1/day) recently reported by Gizis et al. (2017).

Obtaining an estimate of the brightness temperature is difficult since the resolution of our observations does
not provide a precise estimate of the size of the emitting region. Additionally, as said above, the fraction of the radio emission which is originated at each component of the central binary is unknown. Under the assumption that radio emission comes from a single object of size 0.12\,R$_\odot$ (derived from the models of Chabrier et al. 2000), we calculate a brightness temperature of 
5.4$\times$10$^7$\,K ($\times$1/2 for equal binary contribution), which is consistent with synchrotron or gyrosynchrotron non-thermal radio emission (Dulk 1985).  In principle, the low degree of circular polarization (less than 15\%) seems to discard coherent mechanisms predicted for ultracool dwarfs (i.e., auroral emissions; Hallinan et al. 2015; Kao et al. 2016), normally associated to a high degree of polarization; however, in case that both components A and B contribute to the radio emission, 
we notice that the degree of circular polarization above would be the result of the combination of both radio emitters, not reflecting 
properly the polarization properties of each one. 
Further information about the emission mechanism acting on this object can be obtained from the 4\,GHz recorded bandwidth of our $X$-band VLA observations. In practice, we produced four narrower-band images of VHS\,1256-1257 by deconvolving adjacent 1\,GHz-bandwidth data sets separately (see Fig.\,2), from which
 we could obtain an indication of the spectral behaviour of this system between 8 and 12\,GHz. The corresponding spectral index is 
 $\alpha$ = $-$1.1$\pm$0.3 (S $ \propto \nu^{\alpha}$), compatible with optically thin non-thermal synchrotron or gyrosynchrotron emission from a power-law energy distribution of electrons, indicating, in turn,  that strong magnetic fields play an active role in this system.

If the optically thin regime would hold until frequencies as low as 1.4\,GHz, we should have detected radio emission in VHS\,1256--1257 with a flux density above 300\,$\mu$Jy (actually, this was the motivation for the 1.6\,GHz VLBI observations reported in Sect. 2.2); however, no flux above 20\,$\mu$Jy is detected at the nominal position of VHS\,1256-1257 at $L$-band. Ultracool dwarfs have shown to be strongly variable source in radio (Bower et al. 2016), therefore arguments of variability could explain this lack of detection. However, the persistent non-detection in our VLBI observations (with a noise floor 10 times smaller than the expected 1.6\,GHz
flux density according to the spectral index derived) led us to formulate a different hypothesis consisting in considering that the radio emission is actually self-absorbed at the frequency of 1.4\,GHz.

We have further explored this hypothesis following the analytic expressions developed by White et al. (1989; W89) for gyrosynchrotron radio emission of dMe stars in quiescent conditions, which provide estimates of the spectral index for the optically thick/thin components, and, therefore, the turnover frequency. We notice that synchrotron radio emission is not ruled out by our data, but gyrosynchrotron from midly relativistic electrons seems to be the preferred mechanism for previously studied M-dwarfs (i.e., Osten et al. 2006; 2009), which in turn justifies 
the use of W89 formulation. These authors assume a dipolar magnetic field which scales as $B(r)\propto\,r^{-n}$, with $n=3$ and $r$ the distance measured from the dipole, and a power-law electron distribution $N'(E)$ which also scales as $N(E)\propto N'_o(E)\,r^{-m}$, where the index $m$ varies 
from 0 (isotropic electron distribution) to $m=3$ ($=n$, radial dependence of the electron distribution being the same as that of the magnetic field).
Taking our measured optically-thin spectral index  ($\alpha=-1.1\pm0.3$), which implies an energy index 
$\delta=2.6$,  W89 expressions provide two values for the spectral index of the optically thick 
component of the radiation, $\alpha=0.6$, and $\alpha=1.2$, corresponding to the two extreme cases of 
$m=0$, and $m=3$, respectively. With the constraints above, we can set lower bounds to the turnover frequency of 
$\sim$8\,GHz ($m=0$) and $\sim$5\,GHz ($m=3$), effectively limiting the turnover frequency to the range 5 -- 8.5\,GHz. 

In addition,  for gyrosynchrotron emission, 
the turnover frequency depends strongly of the magnetic field (and very weakly of 
the rest of the model parameters, $m$ and $n$ in particular; W89) in the form B$\sim$150\,$\nu^{1.3}$, which provides magnetic field intensities for the previous turnover frequency range of 1.2--2.2\,kG. Interestingly, the previous 
values agree with M-dwarf magnetic field estimates derived from theoretical models (Reiners et al. 2010a): from the luminosity and mass reported for the components of VHS\,1256-1257A and B (Stone et al. 2016; Rich et al. 2016), and using the radius of 0.12\,R$_\odot$  the model of Reiners et al. (2010a) provides values of the dipole field of $\sim$1.7\,kG, well within the range derived previously. This estimate is near the average value of the magnetic field intensity found in a sample of M7--9.5 dwarfs (Reiners et al. 2010b).

\begin{figure}
\includegraphics[trim={0cm 13cm 3.8cm 5cm},clip, width=7.5cm, angle=0]{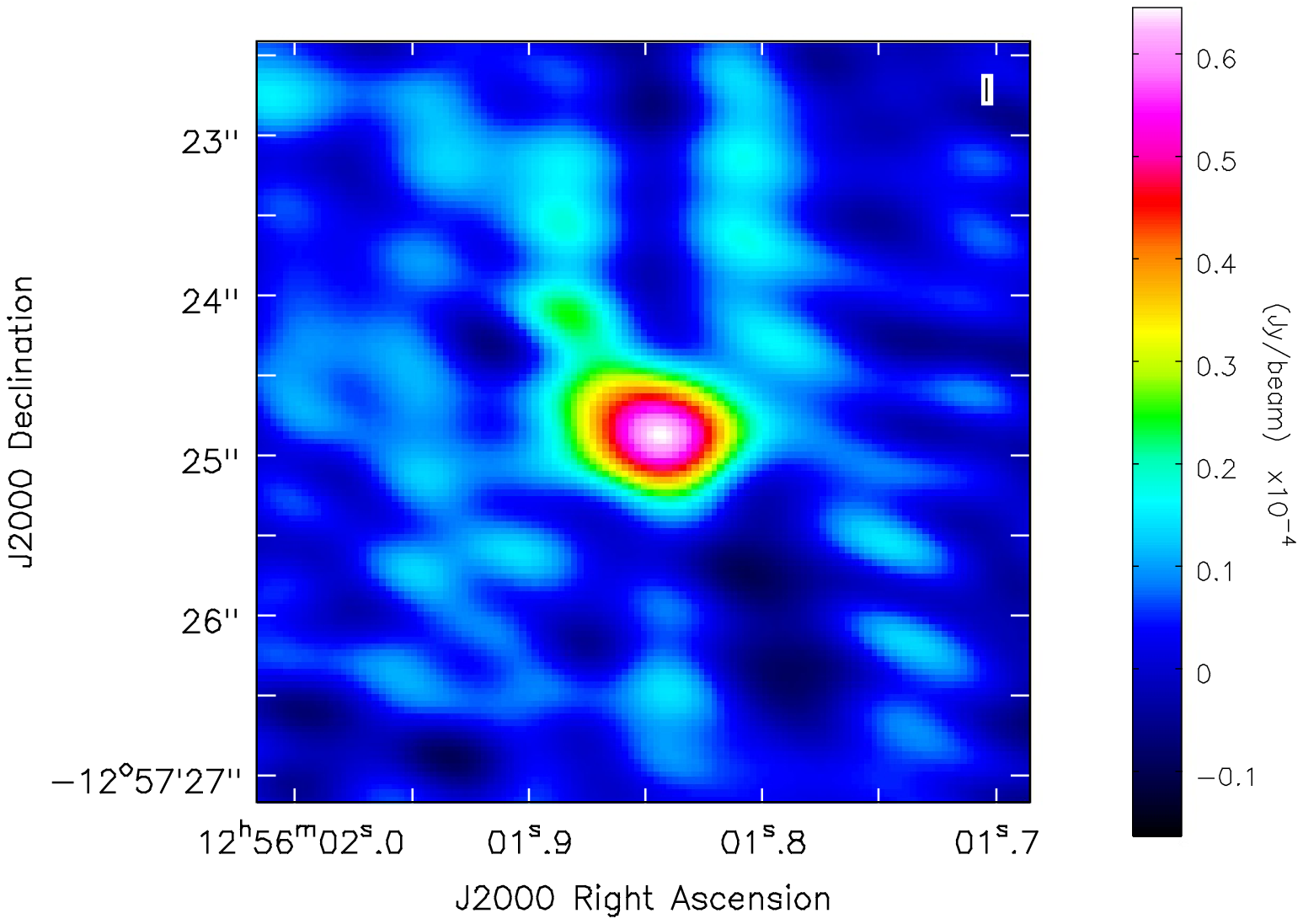}
\includegraphics[trim={0cm 13cm 3.8cm 6cm},clip, width=7.5cm, angle=0]{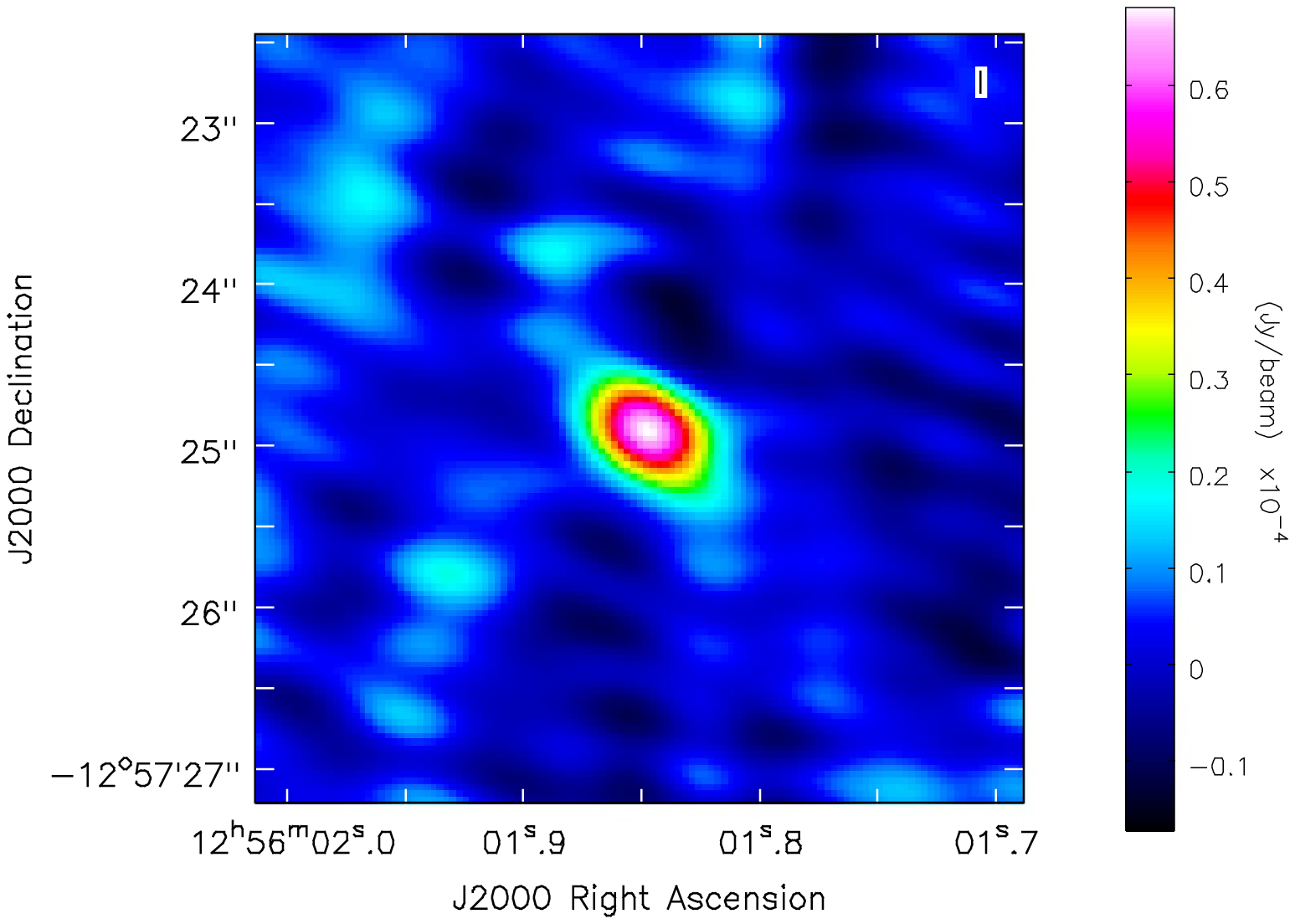}
\includegraphics[trim={0cm 13cm 3.8cm 6cm},clip, width=7.5cm, angle=0]{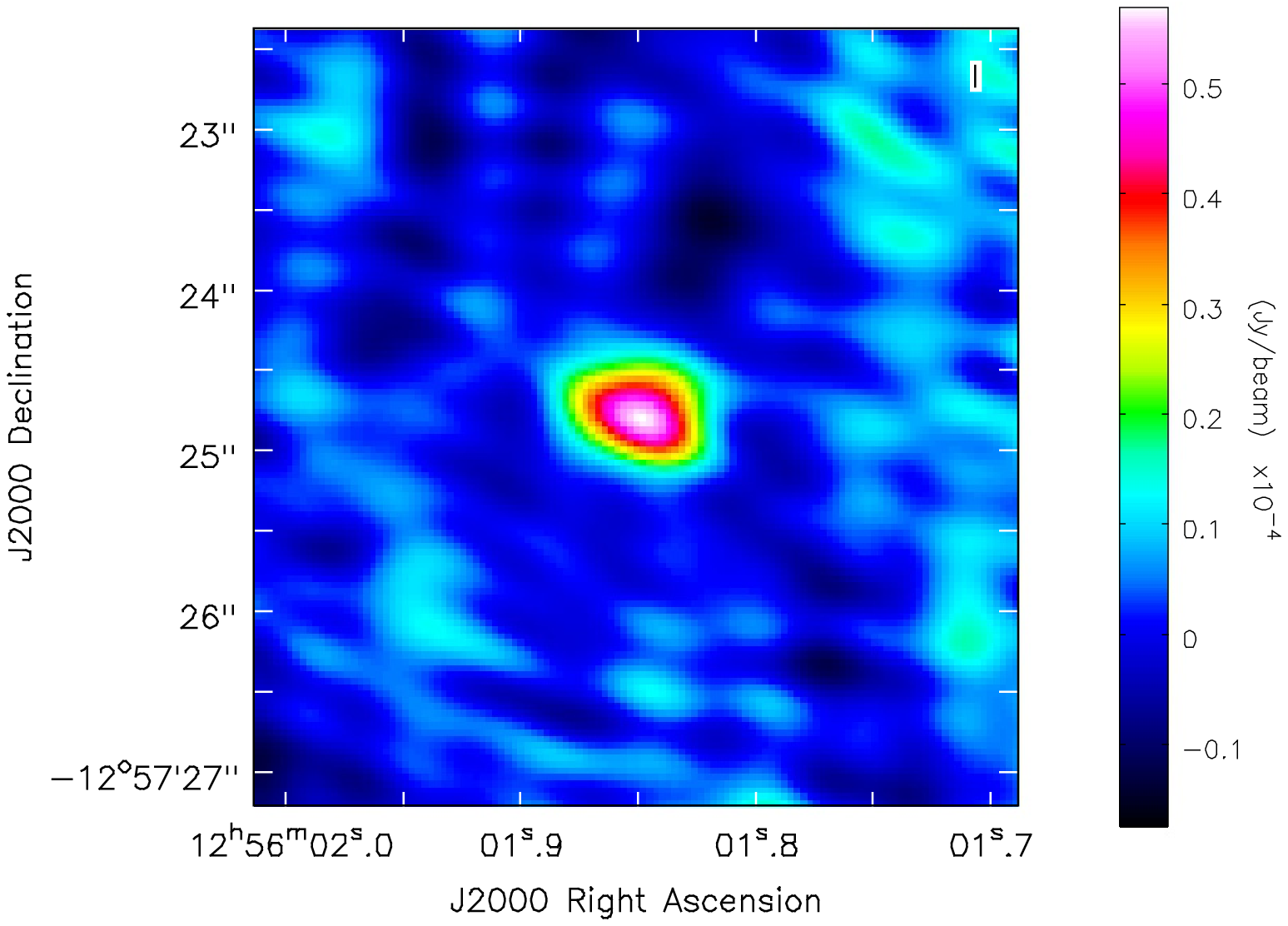}
\includegraphics[trim={0cm 12cm 3.8cm 6cm},clip, width=7.5cm, angle=0]{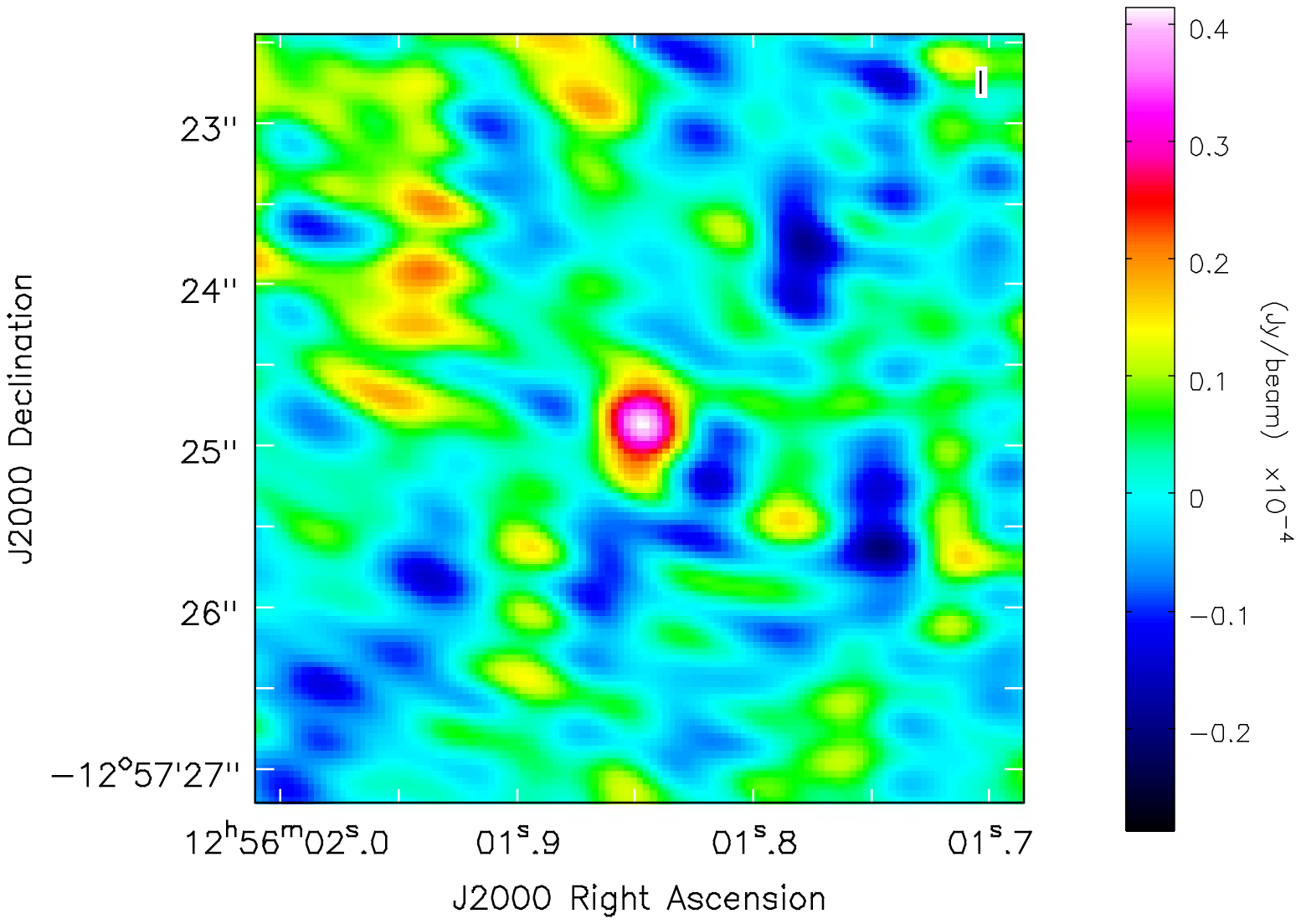}
\caption{VLA images the VHS\,1256$-$1257 system (unresolved central binary AB) at $X$-band made from
1\,GHz subsets of the total data. From top to bottom, the center frequency is 8.5, 9.5, 10.5, and 11.5\,GHz, respectively. The peak flux density decreases from 63\,$\mu$Jy (top) to 45\,$\mu$Jy (bottom image).}

\label{vhsvlal}
\end{figure}

\begin{figure}
\includegraphics[trim={0cm 0cm 0cm 0cm},clip, width=9cm, angle=0]{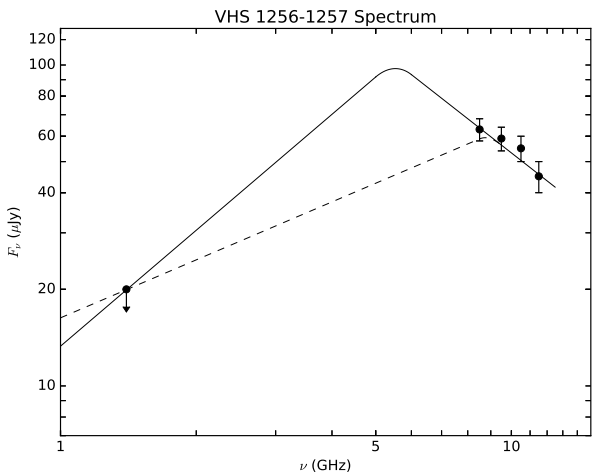}
\caption{Spectrum of VHS\,1256–1257 from VLA observations.
The flux densities between 8 and 12\,GHz
correspond to the maps shown in Fig. 2. The upper
bound resulting from the non-detection at $L$-band is denoted
with a downwards arrow. The solid lines illustrate two possible spectra, each one made from a combination 
of the behaviour predicted by the 
White et al. (1989) model of gyrosynchrotron emission (optically thick regime, showing the two extreme cases, 
 $\alpha$=0.6 --dashed line-- and 1.2 --solid line--), and a fit to our flux density measurements (optically thin regime, 
$\alpha=-1.1$).}

\end{figure}

\subsection{The spectral energy distribution of VHS\,1256--1257\,AB}
\begin{figure}
\includegraphics[trim={0cm 0cm 0cm 0cm},clip, width=9cm, angle=0]{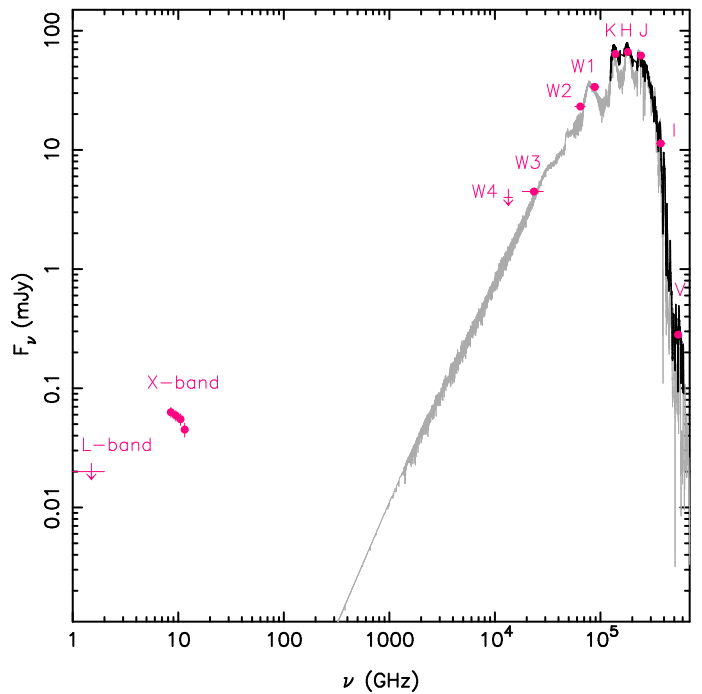}
\caption{Observed spectral energy distribution of the unresolved binary VHS\,1256$-$1257AB from optical wavelengths through 1 
GHz. The optical and near-infrared observed spectra (Gauza et al. 2015) are shown with a solid black line, while the photometric 
observations are plotted as pink symbols: solid circles stand for actual detections with SNR $\ge$ 4 and arrows represent 
4-$\sigma$ upper limits. The filter and passband names are labeled. The horizontal error bars represent the width of the filters. 
Also plotted is the BT-Settl photospheric model (gray line, Baraffe et al. 2015) normalized to the $J$-band flux of 
VHS\,1256$-$1257AB. This model corresponds to a cool dwarf with solar metallicity, $T_{\rm eff}$\,=\,2600 K, and 
log\,$g$\,=\,5.0 [cm s$^{-2}$], which are the parameters expected for an M7.5 source (0.06--0.072 M$_\odot$) with an age of a few 
hundred Myr (Chabrier et al. 2000).}
\label{fig:sed}
\end{figure}

Figure~\ref{fig:sed} shows the complete spectral energy distribution (SED) of VHS\,1256$-$1257AB covering from VLA 
through optical observations. The data at visible, near- and mid-infrared wavelengths are taken from Gauza et al. (2015). 
The observed spectrum is conveniently flux calibrated using the 2MASS $JHK$ magnitudes and the zero points given in 
Cohen et al. (2003). The $W4$ photometry reported in Gauza et al. (2015) is affected by a large uncertainty indicative of 
S/N $\le$ 4. Therefore, we adopt the nominal sensitivity limit of the {\sl WISE} mission at 22 $\mu$m (Wright et al. 2010). 
The BT-Settl solar metallicity model atmosphere (Baraffe et al. 2015) computed for $T_{\rm eff}$\,=\,2600 K and 
log\,$g$\,=\,5.0 [cm s$^{-2}$] is also included in Figure~\ref{fig:sed} to illustrate the expected photospheric emission at 
frequencies not covered by the observations. The synthetical spectrum is normalized to the $J$-band emission of 
VHS\,1256$-$1257AB. This temperature and surface gravity are expected for dwarfs near the star--brown dwarf borderline with an 
age of a few hundred Myr (Chabrier et al. 2000). They also agree with the spectral type---$T_{\rm eff}$ relationship 
defined for high-gravity, ultracool dwarfs by Filippazzo et al. (2015) and Faherty et al. (2016). The BT-Settl photospheric 
model extends from $\sim$300 up to $\sim7.5 \times 10^5$ GHz and does not overlap in the frequency axis with the VLA 
observations. A linear extrapolation of the theoretical SED down to 10 GHz yields a predicted photospheric flux of 
$\sim8.3 \times 10^{-7}$ mJy. The observed VLA $X$-band flux is $\sim$65,800 times higher than the expected photospheric 
emission suggesting that the mechanism responsible for the emission at 10 GHz is extremely powerful. 

\subsection{The radio emission of the very low-mass companion VHS\,1256--1257\,b}

Our non-detection at $X$-band put a strong upper bound to the flux density of this L7 object of  9\,$\mu$Jy (3\,$\sigma$). 
Certainly, the ultracool dwarf samples carried out by  McLean et al. (2012) and Route \& Wolszczan (2013) show that radio detections at GHz-frequencies are not frequent for objects later than L3.5, which could be expected by the declining activity of the cooler atmospheres of these objects. Despite this, auroral radio emission (based on electron cyclotron maser emission mechanism) 
has been detected in a number of late L and T dwarfs  (i.e., Hallinan et al. 2015; Kao et al. 2016; Pineda et al. 2017). Could this emission be expected 
in VHS\,1256-1257\,b? At a distance of 15.8\,pc, the reported luminosities of the coolest dwarfs detected 
in radio (Pineda et al. 2017) would produce a quiescent emission of $\sim$4\,$\mu$Jy 
(below our 3 $\sigma$ detection level). In case of auroral, pulsed emission, the radio flux would significantly 
rise up to peaks of 100 $\mu$Jy, a factor of $\sim$10 above our detection limit in $X$-band. However, 
we do not see such a peak in our data indicating that, at the moment of the VLA observations, 
VHS\,1256$-$1257\,b did not show 
strong levels of auroral activity similar to those seen in other ultracool dwarfs.
 Kao et al. (2016) found a strong correlation between radio aurorae 
and the presence of the H$_\alpha$ line; however, only 7--13\% of the dwarfs with spectral types between L4 and T8 
display H$_\alpha$ emission (Pineda et al. 2016).  Given the absence of H$_\alpha$ in VHS\,1256-1257\,b,  the statistics above do not favor the presence of auroral radio emission in this object. 

On the other hand, considering the $\sim$kG estimate of the magnetic field for VHS\,1256--1257AB, along with the fact that this triple system could have formed from collapse and fragmentation of the same rotating cloud, component b may have also retained high levels of rotation and magnetic field,  which eventually may produce sustainable radio emission, although variable, explaining our non-detection. Indeed, both the model of Reiners et al. (2010a) and the scaling law reported in Christensen et al. (2009) 
(magnetic field $\propto$ energy flux, valid for fully convective, rapidly rotating objects) predict magnetic fields >10$^2$\,G for an object with mass as low as 10--20\,M$_{\rm Jup}$, and effective temperature of 800--1000\,K (Gauza et al. 2015, Rich et al. 2016). 

Additionally, we can estimate the possible radio emission of VHS\,1256--1257\,b from the Nichols et al. (2012) 
model. These authors consider the auroral emission as originating from magnetosphere-ionosphere coupling currents resulting from an angular velocity shear in a fast-rotating magnetized object.
By assuming the fiducial parameters given in Nichols et al. (2012) (corresponding 
to a Jovian-like plasma), a magnetic field of $\sim$2\,kG, and a rotation 
period of $\sim$2\,hr (as extracted from the distribution of brown dwarf 
rotational periods given in Metchev et al. 2015), we find that  
VHS\,1256-157\,b  may present auroral emission with a peak flux 
density of $\sim$100$\mu$Jy, coincident with the estimate above resulting from Pineda et al. (2017) compilation.  However, since the 
currents proposed by Nichols et al. (2012) are created through magnetic
field reconnections,  the cool atmosphere of 
VHS\,1256--157\,b may hamper the existence of auroral emission, as there are evidences
that magnetic reconnections are not allowed or are suppressed
at temperatures below $\sim$1500\,K (Gizis et al. 2017).

\section{Conclusions}
We have reported the detection of radio emission from the VHS\,1256--1257 system. Given the youthness of the system 
($\sim$300\,Myr), its proximity, (15.8\,pc), architecture (a possible triple substellar system), and presence of a very low-mass substellar 
object at 8$^{\prime\prime}$ from the primary, this detection appears 
relevant to study the role of the magnetic field in brown dwarfs. The radio emission is originated at the central system AB, likely consisting in non-thermal synchrotron or gyrosynchrotron emission in presence of kG-intense magnetic field. 
Further monitoring of the system at intermediate frequencies to those presented here should confirm our finding that the turnover frequency of the radiation 
is located between 5 and 8.5\,GHz. The use of interferometers with higher resolution (eMERLIN or EVN at 5--8\,GHz) should discriminate 
if the radio emission originates in one of the components (A or B), in both (A+B),
or perhaps a sort of interaction between them. 
These higher resolution studies will open
the door to a multiepoch astrometric study directed to the determination of the parallax of the system
(modest 5-mas precise positions would result in a 1 pc-precise distance) and, additionally, to precise
estimates of the masses of the internal pair via monitoring of its orbital motion (4.5\,yr period for a face-on orbit).
VHS\,1256–1257\,b  is not seen in our maps; however, despite our non-detection at the level of 9\,$\mu$Jy, $\sim$100\,G magnetic fields 
are expected in this 10--20\,M$_{\rm Jup}$ object, therefore the presence of GHz-radio emission VHS\,1256–1257\,b  should be further explored, as this would provide useful constraints to the emission mechanism in the coolest substellar objects.

\begin{acknowledgements}
We thank Iv\'an Mart\'{\i}-Vidal and Eskil Varenius (Chalmers University of Technology) for helping with the L-band VLA data reduction. The National Radio Astronomy Observatory is a facility of the National Science Foundation operated under cooperative agreement by Associated Universities, Inc. The European VLBI Network is a joint facility of independent European, African, Asian, and North American radio astronomy institutes. Scientific results from data presented in this publication are derived from the EVN project code EG092. 
J.C.G., R.A., and J.B.C.  were partially supported by the Spanish MINECO projects AYA2012-38491-C02-01 and AYA2015-63939-C2-2-P and by the Generalitat Valenciana projects PROMETEO/2009/104 and PROMETEOII/2014/057. M.A.P.T. acknowledges support from the Spanish MINECO through grant AYA2015-63939-C2-1-P.
\end{acknowledgements}

\end{document}